\documentclass[%
reprint,
superscriptaddress,
 amsmath,amssymb,
pra,
]{revtex4-2}

\usepackage{graphicx}
\usepackage{dcolumn}
\usepackage{bm}
\usepackage{siunitx}
\usepackage{lineno}

\begin{document}


\title{Driving electrons at needle tips strongly with quantum light}

 \author{Jonas Heimerl}
\thanks{These authors contributed equally.}
\affiliation{Department of Physics, Friedrich-Alexander-Universität Erlangen-Nürnberg (FAU), Erlangen, Germany}

\author{Andrei Rasputnyi}
\thanks{These authors contributed equally.}
\affiliation{Department of Physics, Friedrich-Alexander-Universität Erlangen-Nürnberg (FAU), Erlangen, Germany}
\affiliation{Max Planck Institute for the Science of Light, Erlangen, Germany}

\author{Jonathan Pölloth}
 \thanks{These authors contributed equally.}
\affiliation{Department of Physics, Friedrich-Alexander-Universität Erlangen-Nürnberg (FAU), Erlangen, Germany}

\author{Stefan Meier}
\affiliation{Department of Physics, Friedrich-Alexander-Universität Erlangen-Nürnberg (FAU), Erlangen, Germany}

\author{Maria Chekhova}
\affiliation{Department of Physics, Friedrich-Alexander-Universität Erlangen-Nürnberg (FAU), Erlangen, Germany}
\affiliation{Max Planck Institute for the Science of Light, Erlangen, Germany}
\affiliation{ECE Department, Technion-Israel Institute of Technology, Haifa 32000, Israel}

\author{Peter Hommelhoff}
\email{jonas.heimerl@fau.de; peter.hommelhoff@fau.de}
\affiliation{Department of Physics, Friedrich-Alexander-Universität Erlangen-Nürnberg (FAU), Erlangen, Germany}
\affiliation{Max Planck Institute for the Science of Light, Erlangen, Germany}
\affiliation{Faculty of Physics, Ludwig-Maximilians-Universität München (LMU), Munich, Germany}

\date{\today}

\begin{abstract}
\textbf{
Attosecond science relies on driving electrons after photoemission with the strong optical field of a laser pulse, representing an intense classical coherent state of light~\cite{Agostini2024, LHuillier2024, Krausz2024}. Bright squeezed vacuum (BSV) is a quantum state of light intense enough to drive strong-field physics~\cite{Iskhakov2012, Gorlach2023,Rasputnyi2024}. However, its mean optical electric field is zero, suggesting that, in a semiclassical view, electrons should not experience strong driving. The question arises if and how this quantum state of light can generate attosecond science signatures in strong-field photoemission. Here we show that the key signatures of strong-field physics - the high energy plateau and the 10-$U_\mathrm{p}$-cut-off - also appear under BSV driving of a needle tip, but only when we post-select electron energy spectra on the individual photon number of each BSV pulse. When averaging over many BSV shots,  we observe broad energy spectra featuring no plateau. This suggests that BSV-driven electrons behave as if driven by an ensemble of coherent states of light. Our findings bridge strong-field physics and quantum optics, offering insights into BSV and other quantum light states. Our work paves the way for electron quantum state engineering and the use of strongly driven electrons as quantum light sensors.
}
\end{abstract}

\maketitle


Investigating how intense ultrashort light pulses interact with matter is at the heart of strong-field and attosecond physics. Over forty years of research has led to an unprecedented understanding of how electrons are driven by a strong optical field, on the attosecond time and sub-nanometer length scale at atoms~\cite{LHuillier2024,Krausz2024,Agostini2024}, and more recently also at metal surfaces~\cite{Bormann2010, Schenk2010,Zherebtsov2011,Kruger2011,Herink2012,Piglosiewicz2013, Dombi2013}. These insights have led to understanding electron dynamics on their natural time scales, with a stunning precision reaching single-digit attoseconds, in atoms, molecules, solids and at the surface of solids~\cite{Niikura2003,Shafir2012,Pedatzur2015,Kim2023,Dienstbier2023}.

According to the three-step model of strong-field physics, electrons tunnel-emitted into an intense optical field can be driven back to the parent matter~\cite{Corkum1993}. There, they may recombine and emit high harmonic radiation or may rescatter elastically to gain higher energy subsequently. This results in the famous plateau, either in the high harmonic spectrum or in the energy spectrum of the electrons~\cite{Ferray1988,Paulus1994}. This all rests on the assumption that the driving field is a classical sine-wave-like optical field, well justified by the large number of photons involved, rendering quantum fluctuations in the amplitude and phase of the light negligible: Already for a small pulse energy of \SI{1}{\nano\joule} the number of involved photons for a near-infrared driving field is of the order $10^{10}$, leading to intensity fluctuations of $\sim0.001$\,\% according to the Poissonian photon statistics of coherent light. For this reason, the common description of the strong-field light-matter interaction neglects the quantum-optical nature of the driving field. 

Two different approaches have recently been proposed to search for quantum optical signatures in strong-field processes, i.e., to explore {\it strong-field quantum optics}~\cite{Bhattacharya2023, CruzRodriguez2024, Stammer2024_2}. In one approach, classical coherent light is used as the driver for strong-field processes that generate non-classical states \cite{Gorlach2020,Stammer2023,Lange2024}. Initial experiments along this line show the generation of non-classical states at the wavelength of the driver by conditioning on the high harmonic generation, maintaining the classical intuition of the electron dynamics~\cite{Tsatrafyllis2017,Lewenstein2021}. Recent studies show quantum correlations between non-perturbative harmonics~\cite{Theidel2024} and predict non-Gaussian states of harmonics and entanglement between harmonics~\cite{Yi2025}.

\begin{figure}
     \centering
     \includegraphics[width=0.85\linewidth]{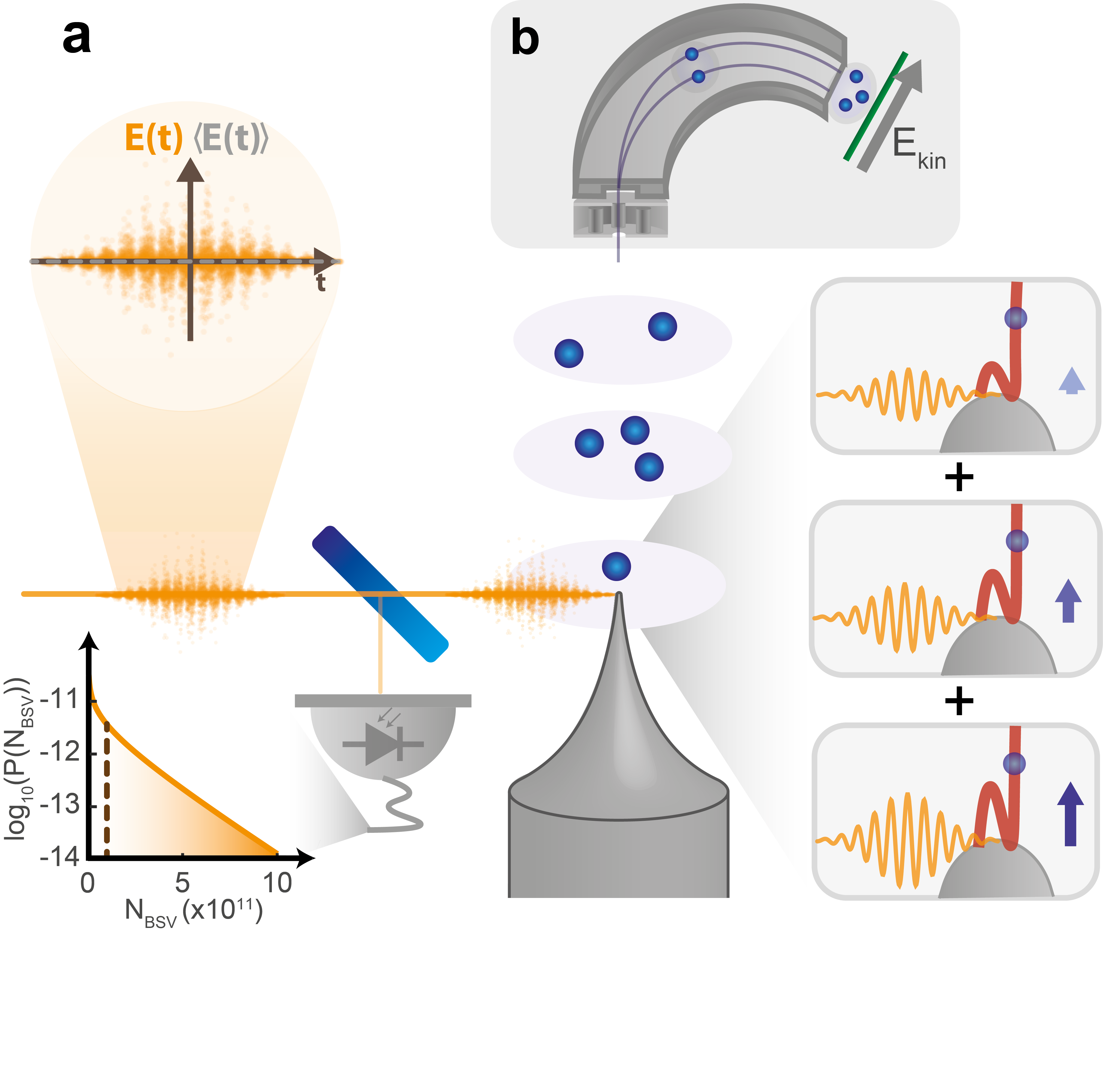}
     \caption{\textbf{Setup for the measurement of strong-field electron energy spectra driven by quantum light.} \textbf{a}, Bright squeezed vacuum (BSV) has a mean electric field of zero ($\langle \boldsymbol E \rangle = 0$, gray dashed line) with a large variance oscillating at twice the carrier frequency (circled inset). 96\,\% of the BSV are sent to the tungsten needle held in an ultrahigh vacuum chamber (not shown), while the photon number of each shot is monitored with the 4\,\% BSV pick-up at a photodiode (gray, losses not included). The calculated photon number distribution $P(N_\mathrm{BSV})$ for a mean of $\langle N_\mathrm{BSV} \rangle = 10^{11}$ photons per pulse (brown dashed line) is shown on the bottom left. Electrons emitted in a non-linear photoemission process are driven in the strong optical near field and rescatter at the tip surface (see boxes), generating high-energy electrons. We discuss below why the insets show classical coherent light states. \textbf{b}, Sketch of the home-built shot-resolving electrostatic low-energy electron spectrometer. We use an electrostatic quadrupole lens in front of the deflector to focus the electron beam emitted from the tip into the spectrometer. The energy of each electron is recorded with the help of a micro-channel plate and phosphor screen placed in the detector plane of the spectrometer (green line).}
     \label{fig:setup}
 \end{figure}

The second approach towards strong-field quantum optics, which we focus on here, is to directly resort to a non-classical driving field of light. We choose bright squeezed vacuum (BSV), an intriguing light state whose mean electric field $\langle \boldsymbol{E}\rangle$ is zero at all times. Its variance oscillates at twice the carrier frequency (Fig.~\ref{fig:setup}a), and it can still represent intense light because its mean intensity $I\propto \langle\boldsymbol{E}^2\rangle$ is proportional to the electric field squared~\cite{Mandel1995}, and $I \propto \sinh^2{r}$, where the squeezing parameter $r$ can be as high as 15~\cite{CHEKHOVA201527,Rasputnyi2024}. Noteworthily, the intuitive semi-classical picture of strong-field dynamics fails dramatically, as no acceleration of an electron would be classically expected since $\langle\boldsymbol{E}\rangle = 0$. Clearly, new theoretical frameworks are needed to take the quantum nature of the light field into account, several of which were recently developed~\cite{Gorlach2020,Fang2023,Gorlach2023,EvenTzur2023,Stammer2023, Wang2023,Even2024, Yi2025, Gothelf2025}.

Initial experiments using non-classical driving light demonstrated that the photon number statistics of BSV is transferred to the number statistics of photoemitted electrons~\cite{Heimerl2024}, and, in a two-color experiment with one color being in a non-classical state, also to the statistics of high harmonic orders~\cite{Lemieux2024, EvenTzur2025}. Furthermore, BSV was used to generate high harmonics in solids~\cite{Rasputnyi2024}. Most recently, a two-color gas-phase high-harmonic generation (HHG) experiment showed in-situ quantum state tomography~\cite{EvenTzur2025}.

Yet, thus far, neither the rescattering of electrons nor the plateau in HHG driven by non-classical light has been demonstrated in experiment. Clearly, such measurements are crucial for verifying current theory models and for investigating how the system of electrons and photons is potentially entangled.
Intriguingly, based on rescattering, the sub-cycle properties of quantum states of light can be probed with attosecond resolution. In addition, such experiments can shed light on the fundamental and non-trivial questions of quantum measurements at extremely short time scales.

\section*{Experiment}
 
In the experiment we generate temporally and spatially single-mode BSV from an unseeded optical parametric amplifier (OPA)  (see Ref.~\cite{Rasputnyi2024} for details and Fig.~\ref{fig:setup}a for the setup). The BSV is centered at \SI{1600}{\nano\meter} (photon energy of \SI{0.77}{\electronvolt}) and has a pulse duration of \SI{25}{\femto\second}. The repetition rate of the pump laser is \SI{1}{\kilo\hertz}. The available pulse energy of BSV is on the order of several hundred nanojoules. We adjust the mean pulse energy for the experiment by varying the pulse energy of the OPA pump. Further, we pick off a small percentage of each BSV pulse by a fused silica window reflection to monitor the pulse-to-pulse photon number fluctuations using a photodiode. The measured photon number at the photodiode $N_\mathrm{BSV}$ is proportional to the number of photons in each BSV pulse. 

The BSV pulses are sent into an ultra-high vacuum chamber with a base pressure of $p < \SI{1e-8}{\hecto\pascal}$, where they are focused to \SI{8}{\micro\meter} ($1/e^2$- intensity radius) using an off-axis parabolic mirror. A metal needle tip with a radius of a few tens of nanometers is situated on a three-axis nanopositioner, with which we align the tip apex to the optical focus. Uncompensated group velocity dispersion leads to temporal broadening of the light pulses to $\sim$$\SI{38}{\femto\second}$ at the tip (still single mode). After the photoemission from the negatively biased tip ($V = \SI{-310}{\volt}$), the electrons travel towards a home-built low-energy spectrometer, which measures both the number of electrons (up to a few tens per shot) and each electron's energy for each individual laser pulse with an energy resolution of $\sim 2$\,eV (Fig.~\ref{fig:setup}b). The electron spectrometer is based on an electrostatic cylindrical deflector analyzer \cite{Yavor2009}. 
A microchannel plate equipped with a phosphor screen is used to image individual electrons with a camera. This camera and the photodiode in front of the vacuum chamber are synchronized to the repetition rate of the laser, allowing us to correlate the photon number $N_\mathrm{BSV}$ and the number and energy of the electrons for each light pulse.

\begin{figure}
   \centering
   \includegraphics[width=0.99\linewidth]{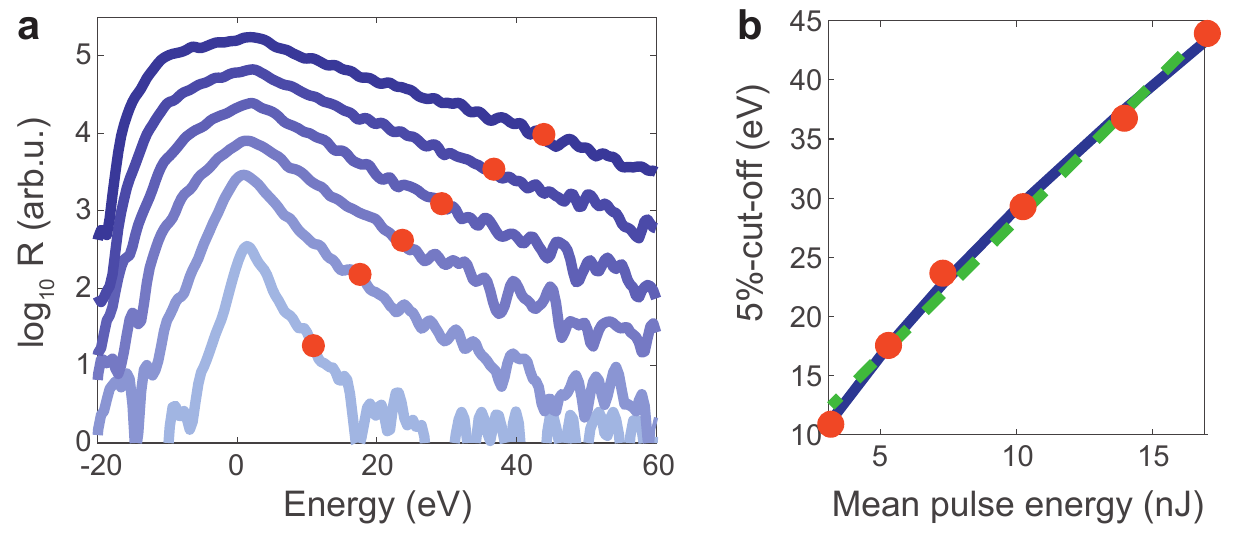}
   \caption{\textbf{Measured shot-averaged electron energy spectra.} \textbf{a}, Electron energy spectra driven by BSV with increasing mean pulse energy (from light blue to dark blue: [$3.2,5.3,7.3,10.2,14.0, 17.0$]\,nJ). For better visibility, the spectra are shifted vertically (the shift of consecutive spectra is 0.4 on the logarithmic axis). The energy offset resulting from the DC bias voltage is subtracted. The red dots mark the points where the count rate has dropped to 5\,\% of each maximum, defining the 5\%-cut-off. \textbf{b}, 5\%-cut-off position as a function of mean pulse energy from panel \textbf{a}. The green dashed line is a linear fit and the blue curve is a power-law fit. The best fit exponent of the latter is $0.65\pm0.2$.}
    \label{fig:Power_scaling_averaged_spectra}
\end{figure}

\section*{Shot-averaged strong-field spectra}

Fig.~\ref{fig:Power_scaling_averaged_spectra}a shows electron energy spectra for a range of mean pulse energies of BSV from \SI{3.2}{\nano\joule} to \SI{17.0}{\nano\joule}. For each spectrum, we record 10,000 images with $\sim$13 laser pulses per image, i.e., each spectrum sums up electrons from a total of $\SI{1.3e5}{}$ BSV pulses.
For increasing pulse energy (light to dark blue) the energy spectra clearly broaden, and the slope around $10...40$\,eV becomes less steep.

Notably, we observe electron energies exceeding \SI{60}{\electronvolt}, where the energy range of the spectrometer ends. For the highest mean pulse energy, we would expect a {\it classical} cut-off energy of \SI{10}{\electronvolt} based on the 10-$U_\mathrm{p}$-law \cite{Paulus1994}, so for the dynamic range of our system, for coherent light electrons with no more than $\sim$18\,eV are expected (beyond the 10-$U_\mathrm{p}$-cut-off; see Extended Data Fig.~\ref{fig:coherent_light_spectra}). The observed energies for BSV exceed this value substantially, but we observe no plateau nor a cut-off.  

To quantify the scaling of the high energy electrons with the mean BSV pulse energy, we define the cut-off for these spectra to lie at 5\% of the maximum rate. For increasing pulse energy, this 5\%-cut-off position shifts to larger values in a slightly sub-linear manner (Fig.~\ref{fig:Power_scaling_averaged_spectra}b). Using a power-law fit to the data (blue), we retrieve an exponent of $0.65\pm0.2$, similar to what is predicted for HHG~\cite{Gorlach2023}. We note that the exact scaling depends on the chosen threshold and also varies slightly from tip to tip.

\begin{figure*}
    \centering
    \includegraphics[width=0.99\linewidth]{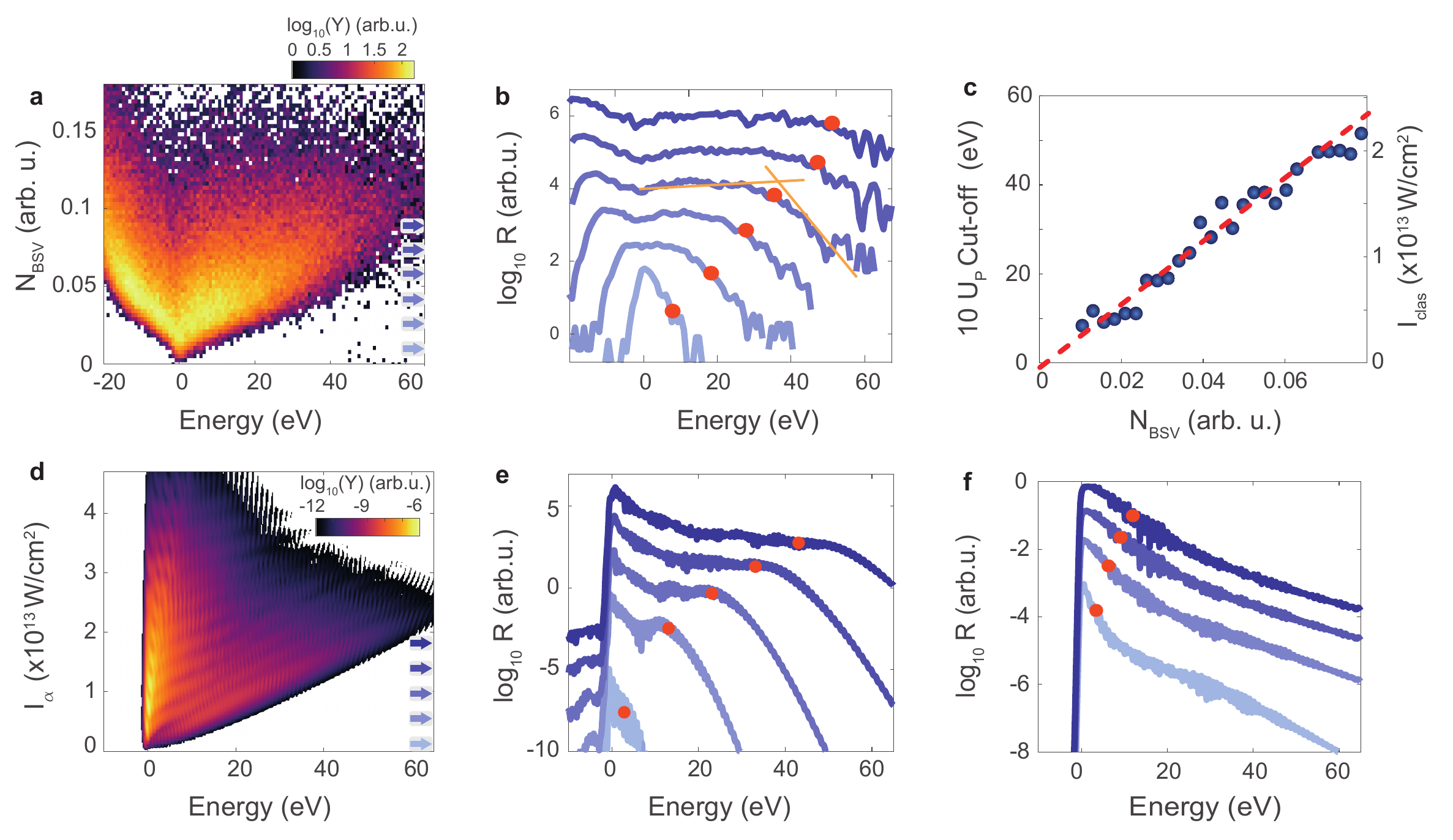}
    \caption{\textbf{Measured and simulated shot-resolved energy spectra.} \textbf{a}, Map of the measured electron energy spectra for a {\it fixed mean BSV pulse energy} of 21\,nJ. The horizontal axis shows the energy of each detected electron and the vertical axis the BSV photon number $N_\mathrm{BSV}$ measured at the photodiode (Fig.~\ref{fig:setup}a). For increasing $N_\mathrm{BSV}$, the energy spectra broaden substantially, from close to the minimum width of 2\,eV up to the maximum detectable electron energy of 65\,eV. 
    \textbf{b}, Six line-out spectra from \textbf{a}. The positions of the lineouts are indicated by blue arrows in \textbf{a}. The shape of these lineout spectra resembles that of well-known strong-driving spectra - with classical coherent laser light: They clearly show the plateau and the 10-$U_\mathrm{p}$-cut-off (red dots). \textbf{c}, Cut-off positions from all line-outs in \textbf{a} as a function of the photon number $N_\mathrm{BSV}$. The right-hand axis indicates the expected intensity $I_\mathrm{clas}$ for coherent light calculated back from the 10-$U_\mathrm{p}$-law (see text). The red dashed line is a linear fit to the data, indicating a scaling behavior identical with coherent driving.   
    \textbf{d}, Simulated shot-resolved electron energy spectra from integrating the time-dependent Schrödinger equation. 
    The vertical axis shows the intensity of a coherent driver $I_\alpha$. The shape matches the experimentally obtained one well, and we stress that these numerical results are obtained with a classical driving field. The fast oscillations are a result of inter- and intra-cycle effects~\cite{Arb2010,Kruger2011}, not visible in the experiment. We note further that we find no electrons at negative energies because the simulation is single electron-based (see Methods). Extended Data Fig.~\ref{fig:Methods_shot_resolved_measurement} contains a version of this figure with more detailed and additional discussions. \textbf{e}, Line-out energy spectra from d for increasing intensities of coherent light ($[0.12, 0.54, 0.96, 1.38,1.8]\times10^{13}\,\mathrm{W/cm^2}$; light blue to dark blue and indicated by blue arrows in d). Red circles indicate the calculated cut-off positions. \textbf{f}, Simulated electron energy spectra for BSV with four mean intensities of $[1.2,2.5, 3.8, 5.0]\times10^{12}\,\mathrm{W/cm^2}$, equivalent to the classical cut-off positions $[3.1, 6.1, 9.0, 12.0]\,\mathrm{eV}$ (indicated by red circles).
    }
    \label{fig:shot_resolved_measurement}
\end{figure*}

\section*{Shot-resolved strong-field spectra}

By our ability to measure individual electron spectra for each light pulse (see Methods), we will now understand the formation of the averaged spectra. Even more importantly, we will conceptually understand how the interaction of BSV and matter in the strong-field regime can be interpreted based on correlating spectra and photon number statistics. Quantum-optically speaking, we will post-select spectra based on the photon number.

We set the mean pulse energy to \SI{21}{\nano\joule} and record \SI{6e5}{} spectra with an average of 0.22 electrons detected per BSV pulse. Electron energy spectra versus the BSV photon number $N_\mathrm{BSV}$  are shown in Fig.~\ref{fig:shot_resolved_measurement}a. Clearly, the electron energy spectra broaden substantially with increasing photon number and exhibit a clear correlation between the width of the spectrum and the number of photons detected in the driving BSV pulse. A dominant broadening is observed towards positive energies, and a smaller one 
to negative. The broadening to negative energies is due to Coulomb repulsion of the electrons. We checked in simulations that Coulomb repulsion effects can be neglected for the large positive energies.

Lineouts of the data in panel a show more clearly the crucial results contained in the spectra-photon number correlation plot (post-selected spectra), shown in panel b.  Intriguingly, the shape of these line-outs resembles those of strong-field spectra well known for coherent driving: a clearly discernible plateau ending in a clearly visible cut-off, which increases with the photon number (see Methods or, e.g., \cite{Kruger2011}). This is a clear indication that the post-selection on the shot's photon number acts like a projection of the driving BSV on the amplitude of a coherent state, for which the semi-classical picture of strong-field physics holds. From this we infer that the clearly visible cut-off is the famous 10-$U_\mathrm{p}$-cut-off. 

To gain quantitative insights, we use two linear functions in the semi-logarithmic representation to extract the 10-$U_\mathrm{p}$-cut-off position for each line-out (red circles), and all BSV bins of Fig.~\ref{fig:shot_resolved_measurement}a \cite{Thomas2013}. The retrieved 10-$U_\mathrm{p}$-cut-off positions are shown in Fig.~\ref{fig:shot_resolved_measurement}c. Identical to coherent driving light, the cut-off position increases proportionally to the post-selected photon number, as shown by a linear fit to the data (red line).  We note in passing that the broad photon-number distribution of BSV in conjunction with post-selection allows us to monitor the electron dynamics without scanning the driving field amplitude.

\section*{Discussion}

The direct connection between photon number and clearly visible 10\,$U_\mathrm{p}$-cut-off allows us to address three important questions: 1) What determines the shape of the {\it shot-averaged} spectra? 2) What is the intensity of the non-classical light averaged over just one optical cycle as experienced by the driven electron? And, 3) does optical field enhancement at the tip apex arise for such a non-classical light field, like for coherent light?

Question 1 is answered by constructing shot-averaged spectra from the shot-resolved spectra.  Adapting pioneering theory work~\cite{Gorlach2023} to electron spectra from metal needle tips, we can obtain an averaged electron spectrum for a fixed mean BSV intensity $ \langle I_\mathrm{BSV} \rangle $ as the sum of individual shot-resolved, i.e. post-selected, electron spectra weighted by the likelihood for this shot's intensity to arise, reflected by the Husimi function of BSV. The Husimi function of BSV is a quasi-one-dimensional distribution $Q_\mathrm{BSV}(\alpha) \approx Q_\mathrm{BSV}(|\alpha|) \delta(\phi_{\alpha})$, where $\alpha$ is the complex amplitude of the electric field and $\phi_\alpha$ its phase (see Methods). The probability distribution of the electric field amplitudes $ Q_\mathrm{BSV}(|\alpha|)$ can be rewritten as function of the (coherent driver) intensity $I_{\alpha}$~\cite{Gorlach2023}, which is equivalent to the expression for the BSV photon number distribution~\cite{Manceau2019}:

\begin{equation}
    Q_\mathrm{BSV}(I_\alpha,\langle I_\mathrm{BSV} \rangle) = \frac{1}{\sqrt{2\pi \langle I_\mathrm{BSV}\rangle I_\alpha}} e^{-I_\alpha/(2\langle I_\mathrm{BSV} \rangle)}.
    \label{eq:husimi_BSV}
\end{equation}

The shot-averaged electron energy spectrum $ S_\mathrm{BSV}(E)$ for BSV driving is then given by
\begin{equation}
    S_\mathrm{BSV}(E,\langle I_\mathrm{BSV} \rangle) = \int  S_\mathrm{coh}(E,I_\alpha) Q_\mathrm{BSV}(I_\alpha,\langle I_\mathrm{BSV} \rangle) \mathrm{d}I_\alpha ,
    \label{eq:spectrum_BSV}
\end{equation}
where $S_\mathrm{coh}(E,I_\alpha)$ is the electron spectrum resulting from driving with classical coherent light with intensity $I_\alpha$ and $E$ is the electron energy (see Methods).

Based on the integration of the time-dependent Schrödinger equation (see Methods), we simulate energy spectra for various {\it coherent} light intensities (Fig.~\ref{fig:shot_resolved_measurement}d, e) and calculate the spectra expected for BSV for four different mean intensities of $[1.2,2.5, 3.8, 5.0]\times10^{12}\,\mathrm{W/cm^2}$ from Eq.~\ref{eq:spectrum_BSV} (Fig.~\ref{fig:shot_resolved_measurement}f). Clearly and like the experimental shot-averaged spectra (Fig.~\ref{fig:Power_scaling_averaged_spectra}a), the resulting simulated spectra feature no prominent plateaus, in contrast to Fig.~\ref{fig:shot_resolved_measurement}e. Further, the sum (see Eq.~\ref{eq:spectrum_BSV}) leads to energies that exceed the classically expected cut-off energies (10-$U_\mathrm{p}$ law, indicated by red dots) by multiples. These shot-averaged spectra, given by the \textit{incoherent} sum, agree qualitatively and quantitatively with the measured ones and directly explain the observed high-energy electrons as well as the washing out of the plateau: Both originate from the interplay of BSV's large intensity fluctuations and in particular intensity outliers and the non-linear electron emission yield (see Extended Data Fig.~\ref{fig:methods_simulation_shot_averaged}).

To address question 2, we consider the driven electrons local probe particles and attain the optical intensities responsible for driving the electrons strongly {\it from the observed cut-off}, which we identified as 10-$U_\mathrm{p}$-cut-off of coherent driving light. We obtain what we call the classical intensity $I_\mathrm{clas} = E_\mathrm{obs-cutoff} \times (c \epsilon_0 m) / (2 e^2) \times (4 \omega^2 /  10.007)$, with $c$ the speed of light, $\epsilon_0$ the vacuum permittivity, $e$ and $m$ the electron charge and mass, $\omega$ the driving light's angular frequency and $E_\mathrm{obs-cutoff}$ the cut-off energy shown in Fig.~\ref{fig:shot_resolved_measurement}b (red spheres). $I_\mathrm{clas}$ is shown on the right-hand axis of Fig.~\ref{fig:shot_resolved_measurement}c. It corresponds to the peak intensity of the strongest optical BSV cycle and includes any potential field enhancement arising at the nanometric needle tip. Hence, we here employ the strongly driven electron as a sensor for the optical electric field of the non-classical light, on a sub-cycle time scale~\cite{Thomas2013} (see discussion on pseudo-coherent state below). 

Furthermore, we obtain the {\it mean BSV intensity experienced by the driven electrons $\langle I_\mathrm{BSV} \rangle$}, i.e., including any potential field enhancement, as $\langle I_\mathrm{BSV} \rangle = \sum_{N_\mathrm{BSV}} Q_\mathrm{meas}(N_\mathrm{BSV}) \times I_\mathrm{clas}(N_\mathrm{BSV}) = \SI{3.9e12}{\watt\per\square\centi\meter}$ for the data in Fig.~\ref{fig:shot_resolved_measurement}, where the measured Husimi function $Q_\mathrm{meas}$ coincides with the photon-number distribution detected at the photodiode (Extended Data Fig.~\ref{fig:Methods_shot_resolved_measurement}). This intensity results from the dynamics of the electron taking place within sub-optical-cycle duration - otherwise we would not expect spectra to look clearly like those from coherently driven electrons.

Question 3 can now be directly addressed: The ratio of this intensity as measured by the electron and the intensity in the bare focus as obtained from the experimental parameters given above yields the square of the field enhancement factor~\cite{Thomas2013}. Similarly but more precisely, we attain the field enhancement factor from the slope in Fig.~\ref{fig:shot_resolved_measurement}c as 
$\mathrm{FE} = 3.4\pm 0.6$. This value agrees with field enhancement factors expected for coherent light and tips of similar size~\cite{Thomas2015}.  Hence it appears that the cut-off energies are proper markers for the peak intensity present at the tip apex and that optical field enhancement also arises for the highly fluctuating BSV, like for coherent light. This will not hold any longer when the build-up time of the field enhancement, i.e., the electron response inside of the tip apex as obtained from the plasma frequency, exceeds the mean fluctuation time of the driving light.

It is essential to realize that we have arrived at a point where we decompose the BSV driving field into a set of classical fields. This is a direct consequence of the measurement procedure based on the post-selection on the particular photon number of the driving field. This post-selection can be seen as a projection onto a pseudo-coherent state, i.e., a coherent state with undefined phase~\cite{Spasibko2017}. In this way, a non-zero optical electric field amplitude results, and we can infer the optical-field statistics through the cut-off energy of the electron resulting from the strong light-matter interaction. Our results do not seem to display contributions of the quantum coherence of the driving light, i.e., that BSV is a \textit{coherent superposition} of coherent states. It will remain for future work to identify observables connected to quantum features of the driving optical field, which may be more easily observable with other quantum states of light, with other post-selection approaches, or with novel photo-electron quantum state tomography techniques~\cite{Laurell2025} applied to strongly driven electrons. Conversely, electrons with energies close to the cut-off can potentially be used for sub-cycle quantum state tomography of light~\cite{Hubenschmid2024}.

\bibliography{bibliography}

\clearpage  

\section*{METHODS}
\setcounter{figure}{0} 
\renewcommand{\figurename}{Extended Data Fig.}

\section{Bright Squeezed Vacuum}

The experimental setup for the generation of single-mode bright squeezed vacuum is described in \cite{Rasputnyi2024}.
The measured second-order correlation function of the light in our experiments is $g^{(2)} > 2.8$, which shows the nearly single-mode character of the generated light. We calculate this value from $Q_\mathrm{meas}$ in Extended Data Fig.~\ref{fig:Methods_shot_resolved_measurement}b (grey curve).\\

In Eq.~\ref{eq:husimi_BSV} and Eq.~\ref{eq:spectrum_BSV} we exploited the fact that the Husimi function of BSV can be simplified to a one-dimensional expression as function of the intensity. More generally, the Husimi function of a squeezed state is given by
\begin{equation}
    Q_\mathrm{BSV}(\alpha=x+iy) = \frac{1}{\pi\cosh(r)}\exp\left(-\frac{2y^2}{1+e^{-2r}}-\frac{2x^2}{1+e^{2r}}\right),
\end{equation}
where $r$ is the squeezing parameter~\cite{Gorlach2023}. For strong squeezing, like in our case, the distribution is heavily stretched along the anti-squeezed quadrature and is close to a delta function along the squeezed one. As shown in Ref.~\cite{Gorlach2023}, one can then approximate the Husimi function by
\begin{equation}
    Q_\mathrm{BSV}(\mathcal{E}_\alpha) = \frac{1}{2\pi|\bar{\mathcal{E}}|^2}\exp\left(-\frac{|\mathcal{E}_\alpha|^2}{2|\bar{\mathcal{E}}|^2}\right)\delta(\phi_{\alpha}),
\end{equation}
where $\mathcal{E_\alpha} \propto \alpha$ is the complex electric field and $\mathcal{E}$ the mean electric field.
Since the intensity is $I_\alpha\propto |\mathcal{E_\alpha}|^2$, we end up with Eq.~\ref{eq:husimi_BSV}, where the additional factor $1/\sqrt{I_\alpha}$ comes from normalization.

\section{Experimental shot-resolved electron energy spectra}

Due to the design of our electron energy spectrometer (see main text), the energy range we can resolve depends on the bias voltage applied to the metal needle tip and the potentials applied to the electron deflector. For a bias voltage of \SI{-310}{\volt}, we can measure up to \SI{60}{\electronvolt} energy gain of the electrons. We can extend this range by two consecutive measurements with two different bias voltages of the tip. This was done for the shot-resolved measurement in Fig.~\ref{fig:shot_resolved_measurement}, with bias voltages \SI{-270}{\volt} and \SI{-310}{\volt}. Via post-processing we merge these two measurements to obtain shot-resolved energy spectra with an extended range.

\section{Experimental strong-field spectra from coherent light}

\begin{figure}
    \centering
    \includegraphics[width=0.85\linewidth]{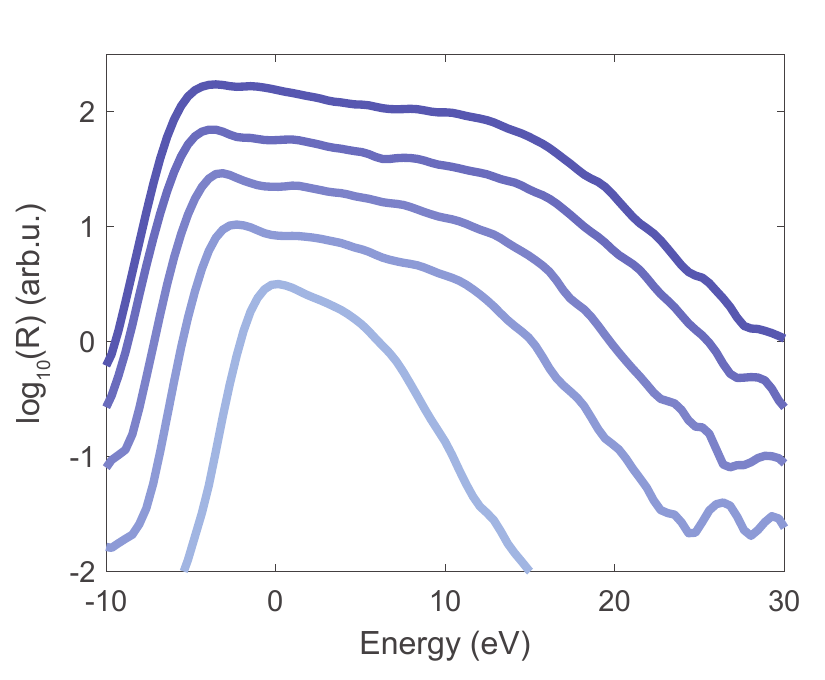}
    \caption{\textbf{Strong-field spectra measured with coherent light at a central wavelength of \SI{1600}{\nano\meter}.} For better visibility, the spectra are shifted on the vertical axis. See text for details.}
    \label{fig:coherent_light_spectra}
\end{figure}

For comparison to the BSV case, we measured averaged strong-field spectra with coherent light pulses at the same wavelength of \SI{1600}{\nano\meter}. In this case, the pulse duration was $\tau = \SI{70}{\femto\second}$. In Extended Data Fig.~\ref{fig:coherent_light_spectra} we show spectra for pulse energies of $[24,33,40,45,51]$\,nJ. We observe a clear plateau for all pulse energies. Furthermore, the maxima of the spectra shift for larger pulse energies towards negative energies, similar to our measurement with BSV in Fig.~\ref{fig:shot_resolved_measurement}a. For the highest pulse energy, where the near field intensity is $I_\mathrm{coh}  = \SI{6.3e12}{\watt\per\square\centi\meter}$, the shift is \SI{3.5}{\electronvolt}. From this we infer that the observed shift of the maxima in the case of BSV driving is not related to the quantum state of the driving light but rather classical Coulomb repulsion of the emitted electrons (see Fig.~\ref{fig:shot_resolved_measurement}b).

\section{Simulation of coherent-light strong-field spectra}

We calculate the energy spectra of the electrons by solving the single-electron time-dependent Schrödinger equation (TDSE) for a particle in a box. For details of the implementation and code examples, we refer to Ref.~\cite{Dienstbier2023}. Here, we assume a work function (ionization potential) of 6\,eV. Further, we include a relatively strong static electric field of \SI{0.5}{\volt\per\nano\meter}, which is present in the experiment due to the bias voltage applied and the small tip size. We assume a pulse duration of $\tau = \SI{25}{\femto\second}$ instead of $\tau = \SI{38}{\femto\second}$, because of computational reasons as the grid size in the simulation scales roughly quadratically with the pulse duration. Both pulse durations represent multi-cycle laser pulses, where the exact pulse duration has no significant effect on the spectral shape. Further, we assume a near-field decay length of \SI{20}{\nano\meter} and a field enhancement factor of $FE =3$, which corresponds to the expected parameters from a tip with few tens of nanometer radius \cite{Thomas2015}.

We simulate the energy spectra for intensities from \SI{1e11}{\watt\per\square\centi\meter} to \SI{5e13}{\watt\per\square\centi\meter} and two CEPs of $\phi_\mathrm{CEP} = [0,1]\cdot \pi$. The spectra for each intensity in the main text is the CEP-average, because of a non-stabilized CEP in the experiment.

We calculate the yield for a fixed intensity by the sum of all energies larger than zero. The simulated yield as a function of the intensity shows, on top of the increase, a periodic modulation (cf. Extended Data Fig.~\ref{fig:methods_simulation_shot_averaged}a). This modulation is due to channel closing.

\section{Simulation of shot-averaged spectra driven by BSV}

\begin{figure}
    \centering\includegraphics[width=0.99\linewidth]{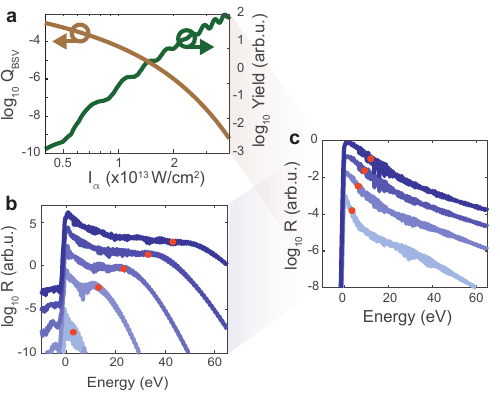}
    \caption{\textbf{Husimi function and simulated shot-averaged electron energy spectra driven by coherent light and BSV.} \textbf{a}, Calculated amplitude part of the Husimi function $Q_\mathrm{BSV}$ (brown) and simulated electron yield (green) as function of peak intensity $I_\alpha$, shown on a double-logarithmic scale. \textbf{b}, Simulated CEP-averaged electron energy spectra for increasing intensities of coherent light ($[0.12, 0.54, 0.96, 1.38,1.8]\times10^{13}\,\mathrm{W/cm^2}$; light blue to dark blue). Red circles indicate the calculated cut-off positions. \textbf{c}, Simulated electron energy spectra for BSV with four mean intensities of $[1.2,2.5, 3.8, 5.0]\times10^{12}\,\mathrm{W/cm^2}$, equivalent to the classical cut-off positions $[3.1, 6.1, 9.0, 12.0]\,\mathrm{eV}$ (indicated by red circles). Panels b and c are identical to Fig.~\ref{fig:shot_resolved_measurement}e, f and are shown here for clarity.}
    \label{fig:methods_simulation_shot_averaged}
\end{figure}

In Extended Data Fig.~\ref{fig:methods_simulation_shot_averaged}a we show the amplitude part of the Husimi function of BSV, $Q_\mathrm{BSV}(|\alpha|)$ (brown curve), calculated for a mean intensity of $\langle I_\mathrm{BSV} \rangle = \SI{1.5e12}{\watt\per\square\centi\meter}$ and, in the same plot, the simulated total yield for coherent driving (green, right axis) as a function of the intensity $I_\alpha$ (see Methods for details of the simulation). Clearly, lower intensities (below $\sim\SI{1e13}{\watt\per\square\centi\meter}$) have a high probability of occurring according to the Husimi function but lead to a small electron yield. Therefore, this contribution to the count rate $R$ in the electron spectrum competes with that from high intensities, which show a higher electron yield but appear less often. 

Based on the integration of the time-dependent Schrödinger equation, we simulate energy spectra for various {\it coherent} light intensities (light blue to dark blue, Extended Data Fig.~\ref{fig:methods_simulation_shot_averaged}b) and calculate the spectra expected for BSV for four different mean intensities of $[1.2,2.5, 3.8, 5.0]\times10^{12}\,\mathrm{W/cm^2}$ from Eq.~\ref{eq:spectrum_BSV} (Extended Data Fig.~\ref{fig:methods_simulation_shot_averaged}c). Clearly and like the experimental ones (Fig.~\ref{fig:Power_scaling_averaged_spectra}a), the resulting spectra feature no prominent plateaus, in contrast to Extended Data Fig.~\ref{fig:methods_simulation_shot_averaged}b. Further, the sum leads to energies that exceed the classically expected cut-off energies (10-$U_\mathrm{p}$ law, indicated by red dots) by multiples. These shot-averaged spectra agree qualitatively with the measured ones and directly explain the observed high-energy electrons as well as the washing out of the plateau: Both originate from the interplay of BSV's large intensity fluctuations and in particular intensity outliers and the non-linear electron emission yield.

\section{Detailed analysis of shot-resolved strong-field spectra driven by BSV}

\begin{figure*}
    \centering
    \includegraphics[width=0.99\linewidth]{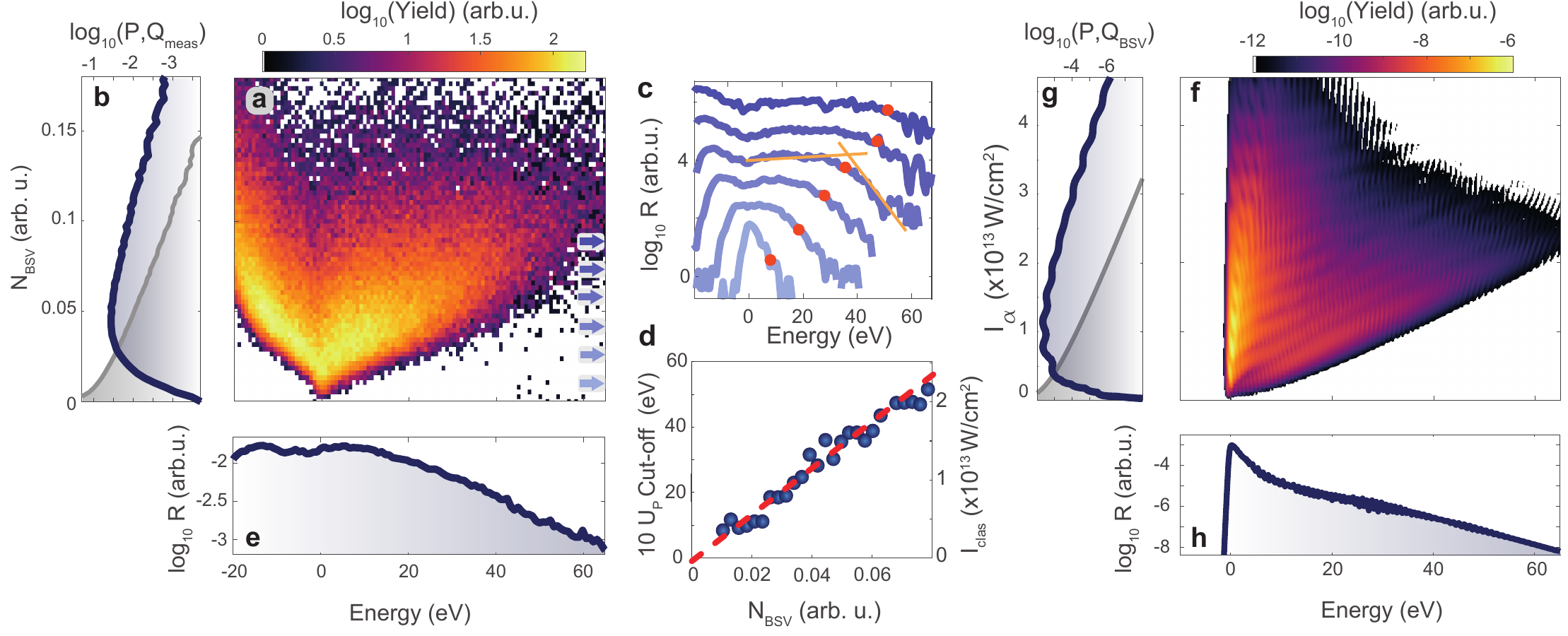}
    \caption{\textbf{Measured and simulated shot-resolved energy spectra.} Extended version of Fig.~\ref{fig:shot_resolved_measurement}. \textbf{a}, Map of the measured electron energy spectra for a {\it fixed mean BSV pulse energy} of 21\,nJ. The horizontal axis shows the energy of each detected electron and the vertical axis the normalized BSV photon number $N_\mathrm{BSV}$ measured at the photodiode. For increasing $N_\mathrm{BSV}$, the energy spectra broaden substantially, from close to the minimum width of 2\,eV up to the maximum detectable electron energy of 65\,eV. \textbf{b}, Normalized yield $P$ of electrons as a function of $N_\mathrm{BSV}$ (blue, marginal distribution of panel \textbf{a} onto vertical axis). This curve is the product of the BSV intensity probability distribution and the total ionization yield (see Fig.~\ref{fig:methods_simulation_shot_averaged}a and Extended Data Fig.~\ref{fig:methods_simulation_shot_averaged}). The gray line is the measured photon number distribution. \textbf{c}, Six line-out spectra from \textbf{a}. These spectra clearly show the plateau and the 10-$U_\mathrm{p}$-cut-off (red dots) as known from coherent driving light. The arrows in \textbf{a} indicate the position of the line out. \textbf{d}, Cut-off positions from all line-outs in \textbf{a} as a function of the photon number $N_\mathrm{BSV}$. The right-hand axis indicates the expected intensity $I_\mathrm{clas}$ for coherent light calculated back from the 10-$U_\mathrm{p}$-law. The red dashed line is a linear fit to the data, indicating also the identical scaling behavior as for coherent driving. \textbf{e}, Shot-averaged energy spectrum obtained by the projection of \textbf{a} on the horizontal axis, matching the shot-averaged spectrum in Fig.~\ref{fig:Power_scaling_averaged_spectra}a.  \textbf{f}, Simulated shot-resolved electron energy spectra from integrating the time-dependent Schrödinger equation and weighting each intensity by the Husimi function (Eq.~\ref{eq:spectrum_BSV}). The vertical axis shows the intensity of a coherent driver $I_\alpha$. \textbf{g,h}, Projections onto the respective axes as in \textbf{b,e}.}
    \label{fig:Methods_shot_resolved_measurement}
\end{figure*}

Here we show more details on the experimental and simulated shot-resolved spectra together with a detailed version of Fig.~\ref{fig:shot_resolved_measurement}, shown in Extended Data Fig.~\ref{fig:Methods_shot_resolved_measurement}. The extended version shows the marginals of the correlation maps, which are the shot-average spectra (e,h) and the probability of measuring an electron event $P$ as a function of the photon number, or intensity (b,g; blue curves). Additionally, we show the measured and calculated amplitude part of the Husimi function,  $Q_\mathrm{meas}$ and $Q_\mathrm{BSV}$, respectively (grey curve in panels b,\,g). 
The mean intensity in the simulated shot-resolved spectra is  $\langle I_\mathrm{BSV} \rangle = \SI{1.5e12}{\watt\per\square\centi\meter}$.

We choose this intensity so that the simulation maximum of $P$ (blue curve) in Extended Data Fig.~\ref{fig:shot_resolved_measurement}g matches that of the experimental data in panel b, again using the calibration via $I_\mathrm{clas}$ for the experimental axis. This intensity deviates from the experimental one obtained above ($\langle I_\mathrm{BSV} \rangle = \SI{3.9e12}{\watt\per\square\centi\meter}$) likely because of the exact shape of the spectra and Coulomb effects. In simulation, we also find that the probability distribution $P$ and in particular its maximum also reflect the interplay of the non-linear electron yield and the fast-decaying Husimi function (cf. Extended Data Fig.~\ref{fig:methods_simulation_shot_averaged}a). Likewise, the simulated correlation map (panel f) shows a similar yield behavior along the vertical axis and, importantly, the same correlation between post-selected photon-number $N_\mathrm{BSV}$ and the width of the electron energy spectra as in the experiment. 

Comparing the simulated and measured shot-resolved electron spectra presented in Extended Data Fig.~\ref{fig:shot_resolved_measurement}a,f, we find that our single-electron theory cannot explain the broadening of the spectra toward energies below zero (plus the bias voltage), which we observe in the experiment (cf. Extended Data Fig.~\ref{fig:shot_resolved_measurement}a,e). Experimentally, we observe a comparable shift towards negative energies for coherent light pulses at similar intensities (see Extended Data Fig.~\ref{fig:coherent_light_spectra}). We expect that Coulomb repulsion of several tens to hundreds of electrons within one light pulse leads to the observed negative energies \cite{Meier2023} (see also Ref.~\cite{Park2012}). This strong repulsion washes out the peak at low energies below \SI{10}{\electronvolt} visible in the simulated shot-averaged spectrum (Extended Data Fig.~\ref{fig:shot_resolved_measurement}h) but does not affect higher energies notably \cite{Meier2023}.

Last we note that the broad range of photon number distribution in the experiment allows us to probe the emission dynamics with BSV over a large range of Keldysh parameters: Intensities exceeding \SI{2e13}{\watt\per\square\centi\meter} are equivalent to a Keldysh-parameter of $\gamma < 0.7$, placing the photoemission deep in the tunneling regime, while the lowest cut-off energies indicate a Keldysh parameter of $\gamma > 2$, in the multi-photon regime. Together with the agreement between the simulated and the measured shot-resolved energy spectra, we conclude that both in the multi-photon and the tunneling regimes the incoherent sum of Husimi function-weighted coherent-light spectra explains our averaged electron spectra.

\subsection*{Acknowledgments} 
We acknowledge Francesco Tani for discussions. This research was supported by the European Research Council (ERC Advanced Grant AccelOnChip), the Gordon and Betty Moore Foundation through Grant GBMF11473, the Deutsche Forschungsgemeinschaft (DFG, German Research Foundation), Project-ID 429529648 (TRR 306 QuCoLiMa) and DFG Project-ID 545591821. J.H.\ acknowledges funding from the Max Planck School of Photonics. A.R.\ acknowledges funding from the International Max Planck Research School for Physics of Light.

\subsection*{Author contributions}
J.H., A.R.\ and J.P. measured the data. A.R.\ built the BSV source, J.P.\ and J.H.\  the experiment. J.H.\ and J.P.\ analyzed the data and J.H., J.P. and S.M.\ performed simulations. J.H., A.R., J.P., M.C.\ and P.H.\ wrote the manuscript. 


\end{document}